\documentclass[aps,prb,twocolumn,amsmath,amssymb,gallery,footinbib,superscriptaddress,floatfix]{revtex4-2}

\def\draftversion{false}

\RequirePackage{ifthen}
\usepackage{graphicx}
\usepackage{dcolumn}
\usepackage{bm}
\usepackage{xcolor}

\ifthenelse{\equal{\draftversion}{true}}{
  \usepackage{showlabels}
  \usepackage[%
]{hyperref}
  \marginparwidth 2.7in
  \marginparsep 0.5in
  \newcounter{comm} 
  
  \def\mlab#1{\marginpar{\small\bf #1}}
  
}{
  \def\dvm#1{}
  \def\cdm#1{}
  \def\msm#1{}
  \def\asm#1{}
  \def\miq#1{}
  \def\mlab#1{}
  
}

\begin{document}

\preprint{APS/123-QED}

\title{Liquid-crystal-like dynamic transition in ferroelectric/dielectric superlattices}

\author{Fernando Gómez-Ortiz}
\email{gomezortizf@unican.es}
\affiliation{Departamento de Ciencias de la Tierra y Física de la Materia Condensada, Universidad de Cantabria, Avenida de los Castros s/n 39005 Santander. Spain.}
\author{Mónica Graf}
\thanks{Present address: Institute of Physics, Czech Academy of Sciences; Prague, 182 21, Czech Republic}
\affiliation{Materials Research and Technology Department, Luxembourg Institute of Science and Technology (LIST), Avenue des Hauts-Fourneaux 5, L-4362 Esch/Alzette, Luxembourg}
\author{Javier Junquera}
\affiliation{Departamento de Ciencias de la Tierra y Física de la Materia Condensada, Universidad de Cantabria, Avenida de los Castros s/n 39005 Santander. Spain.}
\author{Jorge Íñiguez-González}
\affiliation{Materials Research and Technology Department, Luxembourg Institute of Science and Technology (LIST), Avenue des Hauts-Fourneaux 5, L-4362 Esch/Alzette, Luxembourg}
\affiliation{Department of Physics and Materials Science, University of Luxembourg, Rue du Brill 41, L-4422 Belvaux, Luxembourg}
\author{Hugo Aramberri}
\email{hugo.aramberri@list.lu}
\affiliation{Materials Research and Technology Department, Luxembourg Institute of Science and Technology (LIST), Avenue des Hauts-Fourneaux 5, L-4362 Esch/Alzette, Luxembourg}

\date{\today}

\begin{abstract}
Nanostructured ferroelectrics display exotic multidomain configurations resulting from the electrostatic and elastic boundary conditions they are subject to. While the ferroelectric domains appear frozen in experimental images, atomistic second-principles studies suggest that they may become spontaneously mobile upon heating, with the polar order {\sl melting} in a liquid-like fashion. Here we run molecular dynamics simulations of model systems (PbTiO$_3$/SrTiO$_3$ superlattices) to study the unique features of this transformation. Most notably, we find that the multidomain state looses its translational and orientational orders at different temperatures, resembling the behavior of liquid crystals and yielding an intermediate hexatic-like phase. Our simulations reveal the mechanism responsible for the melting and allow us to characterize the stochastic dynamics in the hexatic-like phase: we find evidence that it is thermally activated, with domain reorientation rates that grow from tens of gigahertzs to terahertzs in a narrow temperature window.
\end{abstract}
\maketitle

In ferroelectric/dielectric junctions the development of a homogeneous polarization state is often precluded by the electrostatic penalty associated to the bound charges at the interfaces~\cite{Stepanovich-05,Bousquet-10.2}. A plethora of phases have been observed and predicted in these frustrated systems~\cite{Junquera-23}, such as flux-closure domains~\cite{Tang-15}, polar vortices~\cite{Yadav-16}, skyrmions~\cite{Das-19}, merons~\cite{Wang-20} or the so-called supercrystals~\cite{Stoica-19}.
The stabilization of one of them over the others depends on a delicate balance among elastic, electrostatic and gradient energies, which can be finely tuned with the substrate-imposed strain~\cite{Li-19.2,Dai-23}, the thickness ratio between the ferroelectric and dielectric layers~\cite{Stepanovich-05, Hong-17, Zubko-12, Dawber-05, Dawber-07}, external electric fields~\cite{Nahas20}, and temperature~\cite{Yuan-23}.

The most studied systems in this field are PbTiO$_{3}$/SrTiO$_{3}$ superlattices. At high temperatures, they present thermally disordered local dipoles, resembling a liquid phase with no long-range order~\cite{Zubko16}. 

\begin{center}
  \begin{figure}
     \centering
      \includegraphics[width=\columnwidth]{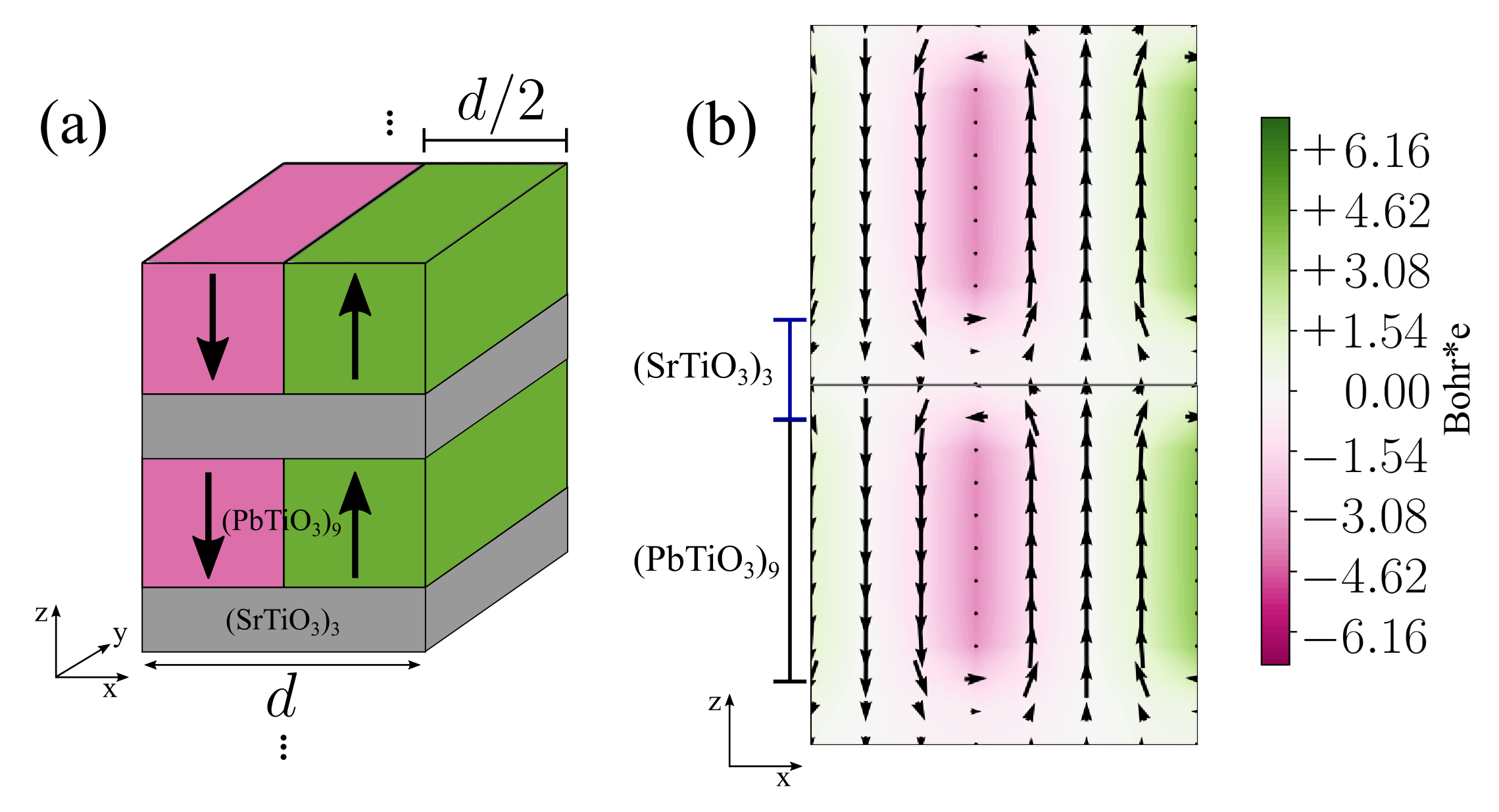}
      \caption{(a) Schematic representation of the (PbTiO$_3$)$_9$/(SrTiO$_3$)$_3$ simulation supercell used for the calculations. The domain periodicity, $d$, amounts to eight unit cells. Green and magenta regions represent positive and negative polarization domains oriented along $z$ as indicated by the solid arrows.
      (b) $(x,z)$ planar view of the local dipole pattern in a cross-section of the infinite stripes. Two replicas of the supercell along $z$ are shown.
      Arrows in the dipole patterns represent the projection of the dipoles onto the $(x,z)$ plane, whereas colors refer to the perpendicular dipole component along $y$.}
      \label{fig:Fig1} 
  \end{figure}
\end{center}

Upon cooling, they undergo a phase transition to a multidomain configuration with regions of opposite polarization in order to reduce uncompensated depolarization fields at the interface~\cite{Meyer-02,DeGuerville-05,Luk-09}. At low-temperatures these domains are well-defined static regions mediated by domain walls~\cite{Catalan-12}, and present long-range order (Fig.~\ref{fig:Fig1}).

Interestingly, at moderate temperatures these domain walls become mobile. The mobility of the domain walls has been previously predicted~\cite{Zubko16,Gomez-Ortiz22,Murillo-23} and is experimentally supported by $x$-ray diffraction data~\cite{Zubko16}. In this intermediate phase the domain walls adopt meandering patterns~\cite{Gomez-Ortiz22,Rusu-22} and present spontaneous stochastic motion. This phase thus lacks long-range translational order in the plane~\cite{Paruch-07}. 

At the same time, experimentally reported structure factors of  PbTiO$_3$ ferroelectric films subject to open-circuit-like electrostatic boundary conditions present four- or two-fold symmetry at high temperatures~\cite{Streiffer-02,Thompson-08}, indicating a possible underlying long-range orientational order (i.e. long-range correlations of domain orientation), even in the absence of long-range translational order.

Intermediate phases with long-range orientational order but no long-range positional order appear in two-step melting processes of two-dimensional systems~\cite{Nelson-1979,Young-1979}, such as in  liquid-crystals~\cite{Lammert-1993}, confined colloidal systems~\cite{terao99}, vortices in superconductors~\cite{guillamon09} or frustrated magnets~\cite{li20}, and even in magnetic skyrmion lattices~\cite{Huang-20}. Is this intermediate phase present also in electrostatically-frustrated ferroelectric films?

In this work we explore this possibility by probing the melting mechanism of the ferroelectric domains in PbTiO$_3$/SrTiO$_3$ superlattices using atomistic molecular dynamics simulations based on second-principles models~\cite{wojdel13,garciafernandez16,escorihuelasayalero17} (see appendix for details on the simulation). 

\begin{center}
  \begin{figure}
     \centering
      \includegraphics[width=\columnwidth]{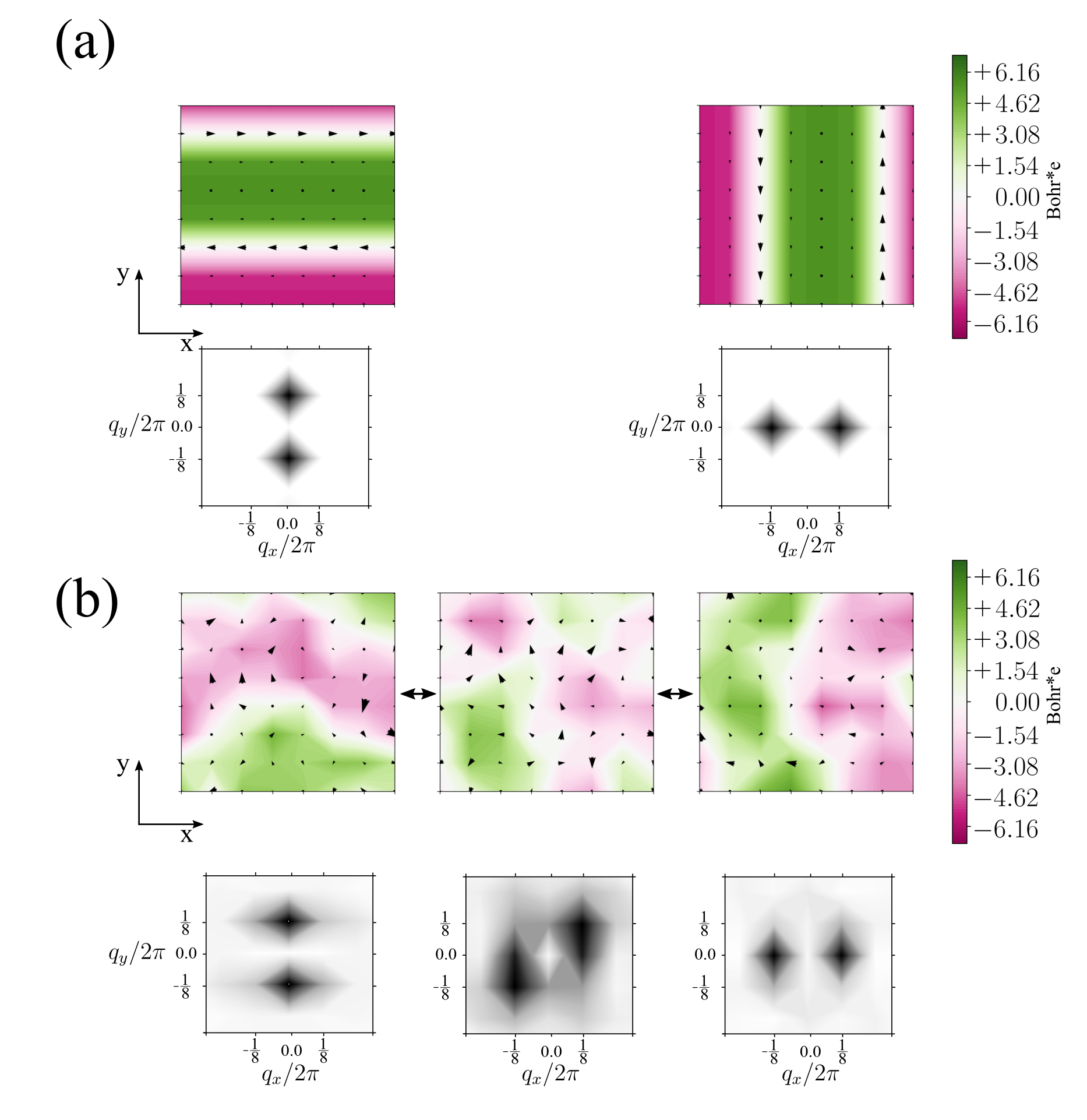}
      \caption{(a) Top row, $(x,y)$ planar view of the local dipole patterns at low temperature ($T=20$~K), where $x$-oriented (left) or $y$-oriented (right) stripe domains are observed depending on the system realization. The arrows at the domain wall are the Bloch component of the polarization. Bottom row, instantaneous structure factors of the $z$-component of the polarization for each of the realizations in units of inverse of the unit cell. (b) Same as in (a) at a higher temperature ($T=280$~K) where we can see (i) a meandering domain wall, and (ii) the motion of domains transitioning periodically from $x$- to $y$-oriented domains (indicated by the double arrows). Arrows in the dipole patterns represent the projection of the dipoles onto the $(x,y)$ plane, whereas colors refer to the perpendicular dipole component along $z$.}
      \label{fig:Fig2} 
  \end{figure}
\end{center}

To this end, we consider the superlattice with 9 layers of PbTiO$_3$ and 3 layers of SrTiO$_3$ --denoted (PbTiO$_3$)$_9$/(SrTiO$_3$)$_3$ in the following-- as a representative model system. We make the in-plane lattice parameter match that of an SrTiO$_3$ substrate, which results in a small epitaxial compression that favors the vertical  direction ($z$) for the dipoles. In agreement with previous works~\cite{Aguado-Puente-12,Zubko16}, we find that at low temperatures the system forms ordered straight stripe domains, where the polarization is essentially pointing upwards or downwards (see Fig.~\ref{fig:Fig1}). At the domain wall, the polarization profiles are closed by the tilting of the local polarization in one perovskite unit cell, as shown in Fig.~\ref{fig:Fig1}(b).
(Note also that a Bloch-like component of the polarization is developed at the domain wall~\cite{Wojdel14a}.)

Importantly, the polarization domains can arrange periodically along any of the symmetry-equivalent directions, $x$ or $y$, as shown in the top row of Fig.~\ref{fig:Fig2}(a). These two possible domain orientations are degenerate in energy. However, as in other spontaneous symmetry-breaking processes, once the system is conformed in a given realization, the structure remains static at low temperatures. 

In order to characterize the dipole structures, we compute their instantaneous structure factors (as described in Eq.~\ref{eq:structurefactor}), which are shown in the second row of Fig.~\ref{fig:Fig2}(a). They show well-defined peaks around $\mathbf{q}=\left(0, \pm 2 \pi/d\right)$~u.c.$^{-1}$ --left--, or $\mathbf{q}=\left(\pm 2 \pi/d, 0\right)$~u.c.$^{-1}$ --right-- (where u.c. stands for perovskite unit cell, and $d=8$ u.c. is the periodicity of the domains), evidencing the fixed periodicity of the stripes and the long-range order of the domains in this phase.

Upon increasing the temperature (280~K $\leq T \leq 320$~K) the long-range ordered (crystalline) phase starts to disorder due to thermal fluctuations. The domain walls become mobile, and the representative microstates correspond to $x$- or $y$-oriented stripe domains that now present meandering walls [see Fig.~\ref{fig:Fig2}(b)].
Moreover, the system starts to transit back and forth between $x$- and $y$-oriented configurations through transient states. A snapshot of a typical transient state is shown in the top central panel of Fig.~\ref{fig:Fig2}(b). 
These configurations retain a vanishing net polarization along $z$, with a more round shape of the domains, showing regions of upwards polarization [green domains in Fig.~\ref{fig:Fig2}(b)] totally surrounded by regions with downwards polarization [magenta domains in Fig.~\ref{fig:Fig2}(b)] or vice versa. Such structures resemble the polarization bubbles that -- according to theory -- can be stabilized in Pb(Zr$_{0.5}$Ti$_{0.5}$)O$_3$ ultra-thin films~\cite{Lai-06} or in PbTiO$_{3}$/SrTiO$_{3}$ superlattices under applied electric field~\cite{aramberri23}, and which have been experimentally observed in Pb(Zr$_{0.2}$Ti$_{0.8}$)O$_3$/SrTiO$_3$/Pb(Zr$_{0.2}$Ti$_{0.8}$)O$_3$ sandwich structures~\cite{Zhang-17}.

The deviations of the domain wall from a straight plane are evidenced by the blurrier peaks in the instantaneous structure factors [see bottom central panel in Fig.~\ref{fig:Fig2}(b)].
 
By further increasing the temperature ($T \geq 330$~K) we find that the typical microstates now resemble the mentioned transient configurations, i.e., the system spends a significant amount of time in such states across the simulation. This high-temperature phase is the liquid phase predicted in Ref.~\cite{Gomez-Ortiz22} (though here it is found at higher temperatures due to the larger ratio between the PbTiO$_{3}$ and the SrTiO$_{3}$ thicknesses).

\begin{center}
  \begin{figure}
     \centering
      \includegraphics[width=\columnwidth]{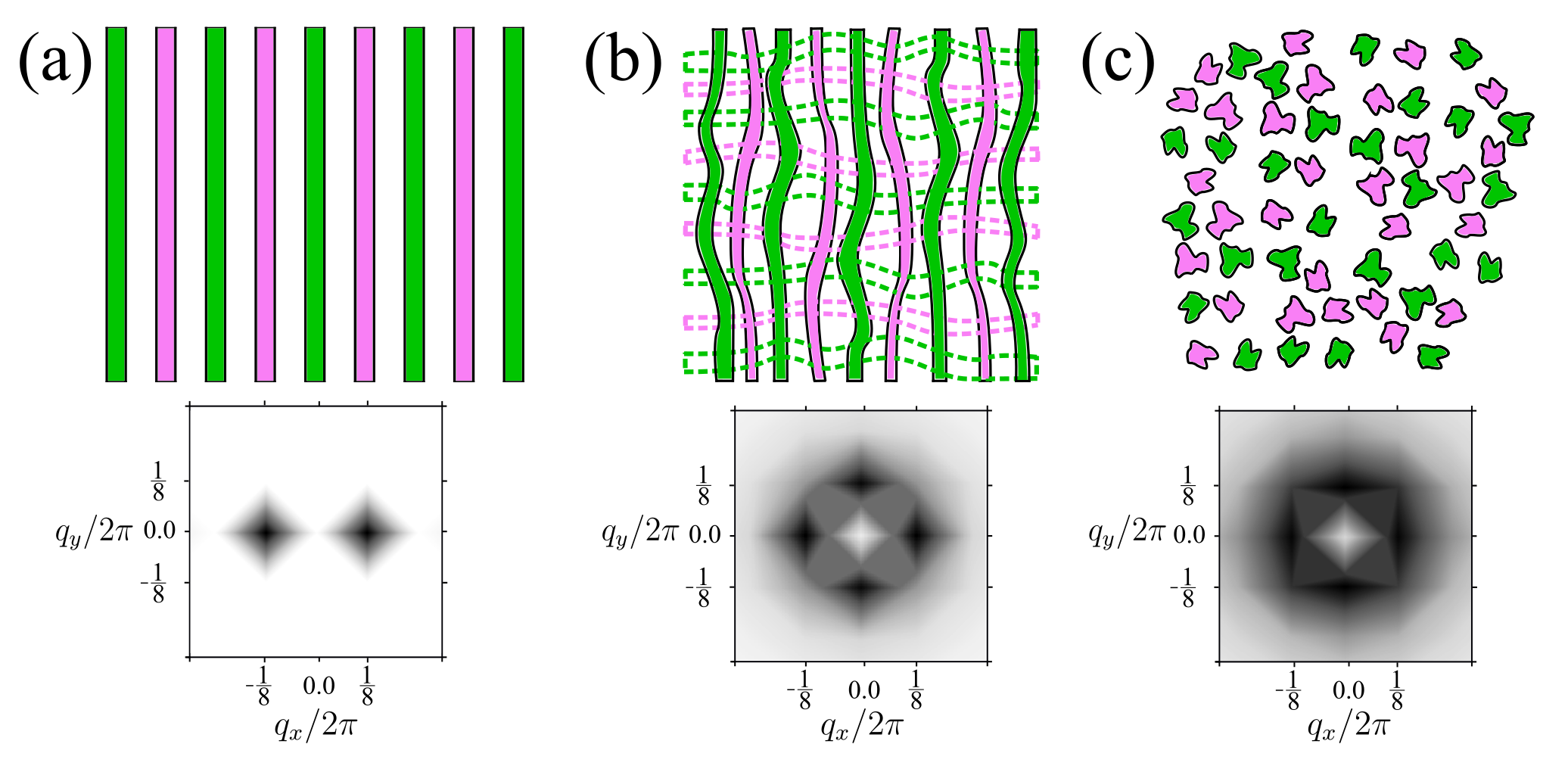}
      \caption{Schematic representation of the different phases adopted by the ferroelectric domains in the PbTiO$_3$/SrTiO$_3$ superlattices upon increasing temperature (upper row), and their corresponding mean structure factor of the $z$ component of the dipoles throughout the whole simulation (bottom row). (a) Low-temperature ($T= 20$~K) crystalline phase characterized by the regular arrangement of stripe domains. Orderings along $x$ or $y$ are degenerate but once the system adopts one configuration it remains static. Only the $y$-configuration is shown.
      (b) Four-fold hexatic-like phase at $T= 290$~K composed of meandering domains flipping between $y$- (solid lines) and $x$-oriented (dashed lines) domains.  The system spends most of the time in these configurations and only a small period of time in transient phases (not shown in the Figure).  
      (c) Smectic-like phase at $T= 340$~K composed of bubble domains that arrange (quasi-)isotropically in plane.}
      \label{fig:Fig3} 
  \end{figure}
\end{center}

In order to characterize the phase sequence, we now resort to statistical analysis of the simulations. In Fig.~\ref{fig:Fig3} we show a qualitative picture of each state (upper row), together with its characteristic mean structure factor of the $z$-component of the dipoles throughout the whole simulation (bottom row). The mean structure factor is computed by averaging the instantaneous structure factor of the configurations (like those displayed in Fig.~\ref{fig:Fig2}) across the corresponding molecular dynamics simulation. (A more complete sequence of mean structure factors as a function of temperature can be found in Supplemental Material~\cite{Supplementary}.)

In the crystalline phase [Fig.~\ref{fig:Fig3}(a)] the mean structure factor peaks along a given direction (either $x$ or $y$ depending on the realization) and shows null intensity elsewhere, indicating that the system is long-lived within this orientation. No distinction can be made between the instantaneous and the average structure factors, suggesting a frozen structure. 

In the intermediate phase [Fig.~\ref{fig:Fig3}(b)] the intensities of $x$- and $y$-oriented stripes become comparable, indicating that the domains spend a similar time along each direction, and only brief periods in the transient phase. The peaks show the same periodicity of domains as in the crystalline phase, but they become slightly blurrier due to the meandering configuration of the domain wall. Moreover, intensities along other directions (e.g. $\lbrace 110\rbrace$) are notably lower. The anisotropic character of the average structure factor reflects that transient states are much shorter-lived and therefore much less probable to occur.

While the size of the simulation box is unfortunately too small to compute long-range correlations explicitly, the intermediate phase clearly shows the defining features of a hexatic phase~\cite{Zaluzhnyy-22}: (i) the long-range translational order is lost as a result of the spontaneous reorientation of the domains and the instantaneous meandering structures, giving rise to four blurry peaks (instead of two sharp ones) in its structure factor; and 
(ii) there is long-range orientational order of the domains, as shown by the four clear peaks in the structure factor, indicating preferred domain orientations. 

Other hexatic phases previously observed show six-fold symmetry (hence the name), and thus a close two-dimensional packing~\cite{Huang-20}. In contrast, the intermediate phase reported here shows a four-fold symmetric structure factor, indicating a strong influence of the perovskite scaffold. Note that crystal structure and polarization are tightly linked in perovskites, while in other systems where hexatic phases have been observed (e.g. superconducting~\cite{guillamon09} or magnetic~\cite{Huang-20} materials) the symmetry of the crystal structure is not so tightly linked with their characteristic order parameter.

Finally, at higher temperatures [Fig.~\ref{fig:Fig3}(c)] the intensities along directions $\lbrace$100$\rbrace$ and $\lbrace$110$\rbrace$ become comparable (with intensity ratios above $0.9$), proving that the system becomes an (almost) isotropic liquid. The (quasi-)isotropic character of the mean structure factor of this phase [bottom row in Fig.~\ref{fig:Fig3}(c)] indicates the loss of long-range orientational order. This is analogous to a smectic liquid-crystal phase. In an early experimental work on ferroelectric PbTiO$_3$ thin films, a transition from four-fold symmetric to random isotropic distributions of stripes as a function of temperature was reported~\cite{Thompson-08}. However, the phase sequence (four-fold to isotropic or vice versa) was shown to strongly depend on the film thickness, and it is unclear whether the observed isotropic character is associated to static (as assumed in their work) or dynamical order (as here predicted).

Note that the emergence of a possible nematic order (where upwards and downwards domains would be able to mix randomly along the normal direction to the layer) is not observed in the simulations, and is probably precluded in ferroelectric ultra-thin layers by construction.

We now focus on the dynamics of the hexatic-like phase, shown in Fig.~\ref{fig:Fig3}(b). 
From the instantaneous structure factors shown in Fig.~\ref{fig:Fig2}b, it is seen how $S(q_{x},q_{y},t)$ clearly differs between the stripe domain configurations arranged along the $x$ or $y$ direction, and the transient configurations.  
For the former the structure factor peaks at $[0,\pm\frac{2\pi}{d}]$ ($[\pm\frac{2\pi}{d},0]$) when the stripes align in the $x$ ($y$) direction. The transient states, by contrast, correspond to either stripe domains along the diagonal or bubble domains in an approximate checkerboard-like tessellation, for which  the bright peaks of the instantaneous structure factor lie along $[\pm\frac{2\pi}{d},\pm\frac{2\pi}{d}]$. 

Tracking the difference in intensities of the instantaneous structure factors allows us to differentiate between $x$- and $y$-oriented stripes by defining the following parameter $I(t)$
\begin{equation}
    I(t)=\left\{ \begin{array}{ll}\dfrac{\vert S(\pm\frac{2\pi}{d},0,t)\vert^2}{\sum\limits_{q_x,q_y}\vert S(q_x,q_y,t)\vert^2} & \mbox{if $S(\pm\frac{2\pi}{d},0,t)>S(0,\pm\frac{2\pi}{d},t)$,} \\-\dfrac{\vert S(0,\pm\frac{2\pi}{d},t)\vert^2}{\sum\limits_{q_x,q_y}\vert S(q_x,q_y,t)\vert^2} & \mbox{if $S(\pm\frac{2\pi}{d},0,t)<S(0,\pm\frac{2\pi}{d},t)$.} \\\end{array}\right.
    \label{eq:intensity}
\end{equation}
\noindent By construction, $I(t)$ is bounded between $-1$ and $1$, and its sign determines whether the stripes are preferentially aligned along $x$ ($I(t) < $ 0) or $y$ ($I(t) >$ 0). 
\begin{center}
  \begin{figure}
     \centering
      \includegraphics[width=\columnwidth]{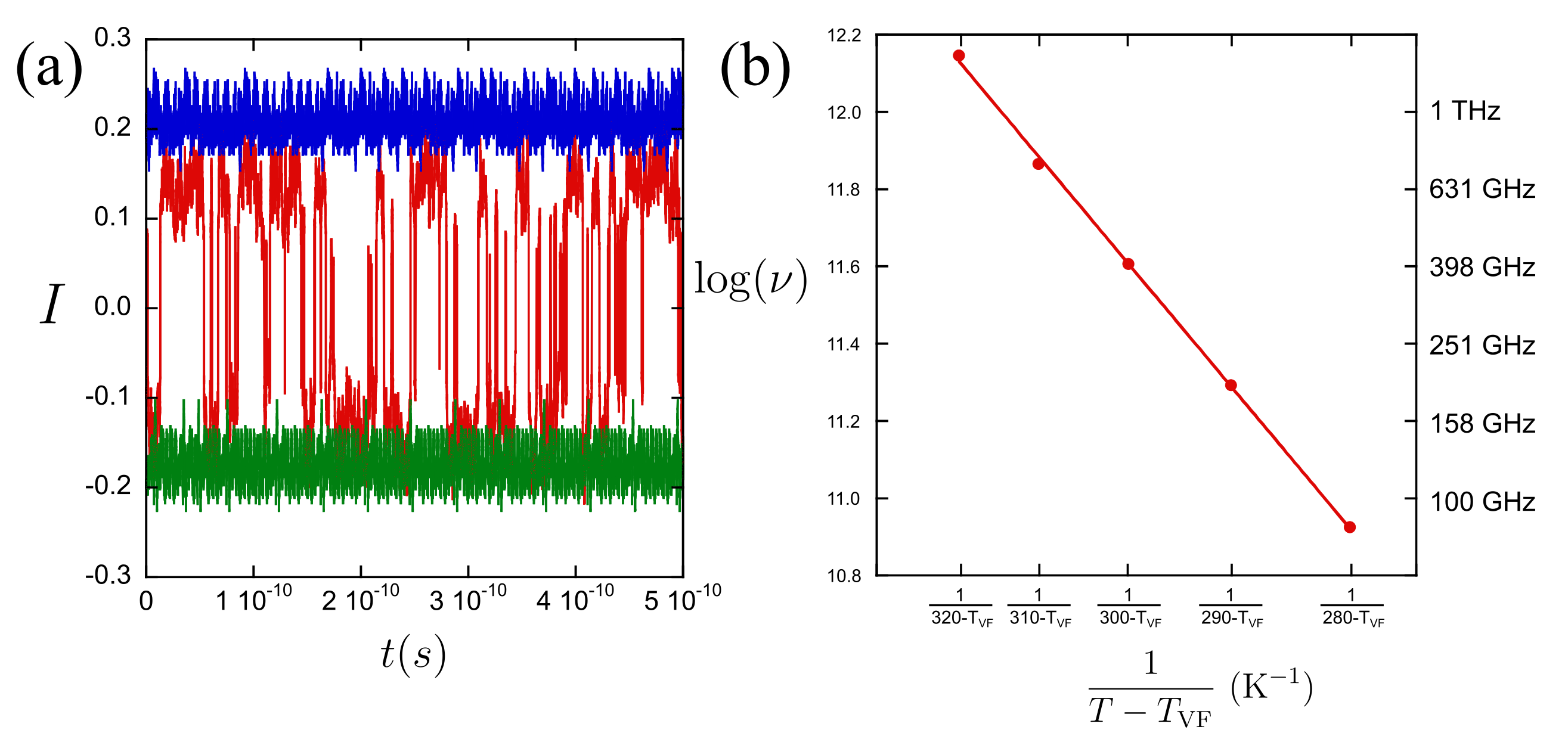}
      \caption{(a) Evolution of the parameter $I(t)$, defined in Eq.~(\ref{eq:intensity}), as a function of time in the molecular dynamics simulations. Green and blue lines correspond, respectively, to different realizations of $x$- and $y$-oriented stripes for the long-lived crystalline-like phase at a temperature of $T=220$~K. The red line corresponds to the hexatic-like phase at $T=280$~K. (b) Decimal logarithm of the mean frequency of hopping between $x$- and $y$-oriented stripes against the inverse temperature showing a Vogel-Fulcher relation, with a Vogel-Fulcher temperature of $T_{\rm VF}=150\pm 40$~K. Right-axis reflects the associated values for the frequency.}
      \label{fig:Fig4} 
  \end{figure}
\end{center}

In Fig.~\ref{fig:Fig4}(a) we show the evolution of $I(t)$ as a function of time in the molecular dynamics simulation. Green and blue data correspond, respectively, to different realizations of $x$- and $y$-oriented stripes for the crystalline phase at a temperature of $T=220$~K. The structure is frozen with $I$ values oscillating around $\pm 0.2$ throughout the whole simulation.
However, when we enter the hexatic-like phase at $T=280$~K (red line) we observe a stochastic oscillating behavior between the $x$- and $y$-orientations. Dividing the number of hoppings by the total time of the molecular dynamics simulation we can estimate the mean frequency of the process, $\nu$. 

By monitoring the evolution of $\nu$ with temperature [Fig.~\ref{fig:Fig4}(b)], we confirm that 
the process is thermally activated, with the frequency ranging from tens of gigahertzs at low-temperatures to terahertzs at higher temperatures. The dependence of the frequency on the temperature can be described by a Vogel-Fulcher relation~\cite{Vogel,Fulcher,Tammann},
\begin{equation}
    \log(\nu)=a+\frac{-E_{\rm VF}}{k_{\rm{B}}(T-T_{\rm VF})},
    \label{eq:VF}
\end{equation}
where $E_{\rm VF}$ and $T_{\rm VF}$ are the Vogel-Fulcher activation energy and temperature, respectively. This model is often used to explain cooperative behavior in supercooled organic liquids~\cite{Adam-04}, spin glasses~\cite{Shrikman-81}, or freezing of polar nanoregions in relaxor ferroelectrics~\cite{Pirc-13}.
The regression of the data to the functional form described in Eq.~(\ref{eq:VF}) [solid line in Fig.~\ref{fig:Fig4}(b)], gives values of $T_{\rm VF}=150\pm40$~K, $a=16.06\pm 0.07$ and $E_{\rm VF}=57.8\pm0.9$~meV. 

At first sight, the obtained value of $T_{\rm{VF}}$ seems to be distant from from the $T=280$~K, the lowest temperature for which we observe domain reorientations. However, taking into account the time scale of our simulations the results can be reconciled. As detailed in the appendix, we use an integration step of $0.5$~fs and the total simulation time is of the order of $10^{-9}$~s. Therefore, we are unable to detect events with frequencies lower than $10^9$~Hz. Extrapolating the Vogel-Fulcher model, we see that such domain reorientation frequencies would occur at a temperature of $T=240\pm 40$~K. This indicates that at this temperature (and down to $T_{\rm VF}=150 \pm 40$~K) the system would not be in the crystalline phase yet; instead it would be slowly hopping between $x$- and $y$- oriented stripes, although in time scales not accessible to our simulations.
Note also that the Vogel-Fulcher temperature is found to lie around 50~K below the melting transition in some materials~\cite{Zhao-12}, suggesting that one should not overinterpret its numerical value.

Near the freezing temperature $T_{\rm VF}$ the characteristic relaxation time for the reorientation of the polar nanoregions is expected to diverge~\cite{Pirc-13}, consistent with our simulations, where we observe long-lived static stripe domains up to $T=270$~K.  
In addition, the activation energy value obtained is consistent with the typical values encountered for relaxor ferroelectrics of around 40-80~meV~\cite{Viehland-90,Pirc-13} and indicates the potential barrier that the stripes have to overcome to start fluctuating.
Note also that Vogel-Fulcher-like dynamics have been reported for closely related ferroelectric/paraelectric superlattices (BaTiO$_3$/SrTiO$_3$) with experimental activation barriers estimated to be between 24 and 53 meV~\cite{lupi23}. The origin of such response was ascribed to relaxor-like behavior associated to shape variations in the dipolar configurations, which might also be the case for our superlattices.

Admittedly, finite-size effects caused by a relatively small simulation box are most probably at play here. We choose the simulation box (with 8$\times$8 perovskite cells in the plane) so that it can accommodate the ground state of the system at $T=0$~K [$x$- or $y$-oriented stripes with a width of 4 perovskite cells, see Fig.~\ref{fig:Fig1}(b)], which is already computationally demanding for molecular dynamics calculations of the length required here. As a result, domain orientations (including stripe domains along alternating $x$ and $y$ orientations, as those observed experimentally in~\cite{das23}) and transition paths that do not fit our simulation cell are not considered in our simulations, which results in an overestimation of the activation energy. Therefore, the actual domain reorientation frequencies should be somewhat smaller than the ones we compute. Nevertheless, we believe our physical analysis remains valid.

In summary, we predict that the polarization domains in PbTiO$_3$/SrTiO$_3$ superlattices undergo a two-step melting process. The intermediate phase bears a strong resemblance to hexatic liquid crystals, showing dynamical long-range orientational order but no long-range translational order. In contrast with liquid crystals, the underlying symmetry of the atomic lattice precludes a six-fold coordination of the stripes, resulting in a four-fold symmetric dynamical structure.
We follow the dynamics of the domain melting and characterize its behavior as a function of temperature, finding a thermally activated domain reorientation of the Vogel-Fulcher type. We show that in a narrow temperature window of $40$~K the domain reorientation rates range between tens of gigahertzs to terahertzs.
Finally, the physical ingredients responsible for the predicted effects, (i) electrostatic frustration precluding the emergence of homogeneous states, and (ii) nanometric domains small enough to be switched by temperature fluctuations, are common to many thin ferroelectrics. Hence, we expect similar two-step melting processes of the polarization domains in other electrostatically frustrated ferroelectric thin films.

\acknowledgments
This work was carried out during F.G-O.'s visit to the Luxembourg Institute of Science and Technology (LIST). 
F.G-O. and J.J. acknowledge financial support from grant No. PID2022-139776NB-C63 funded by MCIN/AEI/10.13039/501100011033 and by ERDF ``A way of making Europe'' by the European Union.
F.G-O. acknowledges financial support from grant FPU18/04661 funded by MCIN/AEI/ 10.13039/501100011033.
H.A. and J.Í-G. were funded by the Luxembourg National Research Fund through Grants C18/MS/12705883 REFOX and C21/MS/15799044/FERRODYNAMICS. We would also like to thank Gustau Catalan from the Catalan Institute of Nanoscience and Nanotechnology (ICN2) for the discussion about thermally activated processes.

\appendix
\section{Simulation methods}
We simulate (PbTiO$_3$)$_9$/(SrTiO$_3$)$_3$ superlattices by means of second-principles methods as implemented in the \textsc{SCALE-UP}
package~\cite{wojdel13,garciafernandez16,escorihuelasayalero17}. Superlattice models were derived from bulk models of PbTiO$_3$ and SrTiO$_3$ developed in previous works~\cite{wojdel13,Wojdel14a}, and have been described elsewhere~\cite{Zubko16,aramberri22}.

We use periodic boundary conditions on simulation boxes of $8\times 8$ elementary perovskite unit cells in the plane and one superlattice period along the stacking direction. For the mechanical boundary conditions, we assume that the in-plane lattice constant is that of a SrTiO$_{3}$ substrate (3.905~\AA). We keep this value fixed throughout all the calculations.

The simulations are carried out following a four-step procedure.
In the first step, performed only once throughout the entire process, we find the ground state of the superlattice.
This is obtained via a Monte Carlo simulated annealing of the system. The simulation begins at 300~K, and the temperature is reduced by a factor of 0.9975 at each step for a total of 20,000 steps. We note that the exact value of the initial temperature at this step does not play a significant role on the final configuration obtained.
The initial atomic positions were chosen to mimic a pair of Bloch-like domains, where the polarization lies parallel to the 180$^\circ$ domain wall and is opposite for neighboring walls. 
The initial stripes can be aligned with the $x$- or $y$-direction, being both degenerate in energy [Fig.~\ref{fig:Fig2}(a)]. 

The second step is the thermalization of the structure at fixed given temperatures. 
For every temperature considered in this work, we run a Monte
Carlo simulation for a total of 50,000 sweeps, always using the ground state (found with the simulated annealing described above) as the starting configuration (and keeping the in-plane strain fixed to that of a SrTiO$_3$ substrate). The initial 10,000 sweeps are considered thermalization steps. We use the remaining 40,000 sweeps to obtain the average out-of-plane equilibrium strains.

The goal of the third step is to obtain a thermalized distribution of atomic velocities for molecular dynamics simulations.  
Fixing the strains to the equilibrium values obtained at the end of the second step, we run a molecular dynamics simulation in the canonical (NVT) ensemble, in which we rescale the velocities every 100 steps. To this end, we take the
final atomic structure obtained with Monte Carlo as the initial
configuration. This simulation is run for a total of 20~ps
with a time step of 0.5~fs. In order to avoid the ``flying ice-cube effect''~\cite{harvey98} 
we remove the velocity of the center of mass every 100 steps (50~fs).

Finally, the fourth step is a microcanonical (NVE) molecular dynamics simulation starting from the (velocity-rescaled) configuration obtained in the previous step. In this simulation we use the velocity Verlet algorithm with a time step of 0.5~fs and for a total of 2~ns. The statistical analysis described in this article is based on these NVE simulations only. 

After collecting the geometries of the molecular dynamics simulation in the microcanonical ensemble, we compute the instantaneous structure factors, $S(q_x,q_y,t)$, defined as
\begin{equation}
S(q_x,q_y,t) = \left\vert\sum\limits_{x=0}^{d-1}\sum\limits_{y=0}^{d-1}e^{-i2\pi (x q_x+y q_y)}\bar{P_z}(x,y,t)\right\vert^2,
\label{eq:structurefactor}
\end{equation}

\noindent which is the squared absolute value of the Fourier transform of the $z$-component of the instantaneous polarization averaged over the three central layers of PbTiO$_3$, $\bar{P_z}(x,y,t)$.

Note that for each temperature studied in the main text we repeated steps two to four.
\end{document}